\documentclass[aps,prb,citeautoscript,twocolumn,longbibliography,superscriptaddress]{revtex4-1}
\usepackage[T1]{fontenc}
\usepackage{latexsym}
\usepackage{graphicx} 
\usepackage{epstopdf}
\usepackage{amsmath}  
\usepackage{amssymb}
\usepackage{comment}
\usepackage{cases}
\usepackage{lipsum}
\usepackage{tikz}
\usepackage{makecell}

\usepackage{comment}

\usepackage{soul}
\usepackage{color}
\usepackage[breaklinks=true]{hyperref}
\usepackage{xcolor}

\begin{document}

\title{Correlation-assisted topological and metamagnetic transitions in Rashba-coupled superconductors: $t$-$J$-$U$ model study}

  \author{Tushar Dey}
  \email{tushar.dey@doctoral.uj.edu.pl}
  \affiliation{Institute of Theoretical Physics, Jagiellonian University, ul. {\L}ojasiewicza 11, 30-348 Krak{\'o}w, Poland }
  \author{Maciej Fidrysiak}
  \email{maciej.fidrysiak@uj.edu.pl}
  \affiliation{Institute of Theoretical Physics, Jagiellonian University, ul. {\L}ojasiewicza 11, 30-348 Krak{\'o}w, Poland }
  \author{J{\'o}zef Spa{\l}ek}
  \email{jozef.spalek@uj.edu.pl}  
  \affiliation{Institute of Theoretical Physics, Jagiellonian University, ul. {\L}ojasiewicza 11, 30-348 Krak{\'o}w, Poland }

\begin{abstract}
  Unconventional superconductivity commonly emerges in systems characterized by strong electronic correlations, with layered copper-oxides serving as a canonical example. Observation of spin-momentum locking within Bi-family of the cuprates, compatible with the presence of non-negligible Rashba-type spin-orbit coupling (RSOC), calls for an investigation of the joint effects of electronic correlations and RSOC on pairing in copper-oxide and related superconductors. Employing statistically-consistent variational approximation (SGA), we carry out such an analysis by constructing the phase diagram for the case of square-lattice \textit{t-J-U} model incorporating RSOC. We also investigate the effects of time-reversal-symmetry breaking by Zeeman field, as well as characterize emergent topological superconducting (TSC) states. Chern number $C = \pm 4$ TSC is found in a broad regime of on-site Coulomb repulsion close to half-filling. The latter is not governed by correlations and emerges also within the weak-coupling Bogoliubov-de Gennes (BdG) scheme. Yet, we identify a distinct $C = \pm 2$ TSC state that is driven specifically by electronic correlations via a topological transition occurring with no bulk quasiparticle gap closure, and is accompanied by discontinuous metamagnetic and Lifshitz transitions. Moreover, a qualitatively distinct doping evolution of the $d$- and $p$-wave components of the underlying mixed-parity SC order parameter above the metal-to-insulator transition is demonstrated. Our work points toward the relevance of joint correlation and RSOC effects beyond BdG scheme to phase diagrams of RSOC-coupled superconductors.
\end{abstract}

\maketitle

\section{Introduction}
\label{section:introduction}

Strong local correlations in lattice electron systems support formation of superconducting (SC) states characterized by a breakdown of symmetries beyond those involved in conventional phonon-driven pairing \cite{Spalek2022}. Layered copper oxides, hosting robust $d$-wave high-temperature (high-$T_c$) SC in the vicinity of metal-to-insulator transition, remain among the most extensively studied realizations of this scenario. However, strong correlations are not the only route toward unconventional SC, with spin-orbit coupling (SOC) providing a noteworthy extension of this canonical scenario. Relevance of SOC to SC pairing has been also invoked in the context of materials varying in the electron-electron interaction magnitude \cite{Smidman2017}, including weakly-correlated superconductors, such as those based on rhenium \cite{Strohmeier2026}. In the context of high-$T_c$ SC, SOC in the cuprates and related strongly-correlated materials has received lesser attention, since its impact on low-energy electronic structure has been long considered negligible \cite{Koshibae1993}. Recent spin- and angle-resolved photoemission spectroscopy (SARPES) experiments \cite{Gotlieb2018, Iwasawa2023, Luo2024} challenge this view by providing evidence for wave-vector dependent spin-polarization in representatives of Bi-family of cuprates, Bi$_2$Sr$_2$CuO$_{6+x}$ (Bi2201) and $\mathrm{Bi_2Sr_2CaCu_2O_{8+\delta}}$ (Bi2212), which is compatible with the presence of Rashba-type SOC (RSOC) of magnitude exceeding $10\,\mathrm{meV}$. A plausible microscopic origin of RSOC include inversion symmetry breakdown within individual CuO$_2$ planes due to local structural distortions \cite{Gotlieb2018, Luo2024}, as also discussed earlier in the context of  YBa$_2$Cu$_3$O$_{6+x}$ \cite{Atkinson2020}. Signatures of spin-orbit coupling in the cuprates have been also reported in recent magnetotransport measurements \cite{Barrera2026}. 

The presence of RSOC affects the dominant spin-singlet $d$-wave SC in several key aspects. First, the SC order parameter acquires an admixture of spin-triplet component. Second, the resultant mixed-parity SC may be associated with an appearance of topological superconducting (TSC) state. This, however, requires further adjustments as point nodes present in the quasiparticle spectrum, inherited from the underlying $d$-wave superconductor, remain protected by time-reversal symmetry (TRS), even if the inversion symmetry is broken \cite{Beri2010}. A fully gapped, plausibly topological SC state characterized by Chern number, $C$, may be achieved by application of external magnetic field that breaks TRS explicitly. The latter route toward strong TSC in RSOC-coupled superconductors is supported by analysis based on the weak-coupling Bogoliubov de Gennes (BdG) \cite{Sato2009, Sato2010, Yoshida2016, Daido2016, Daido2017}. For the case of pairing with the dominant $d$-wave SC component, $C = \pm 4$ TSC has been reported close to half-filling. Subsequent work on extended $t$-$J$ \cite{Nally2024} and Hubbard \cite{Lu2018} models in the strong-interaction regime leads to consistent results. The qualitative agreement between weak- and strong-coupling topological phase diagrams suggests that $C = \pm 4$ TSC is driven primarily by single-particle electronic structure, while the role of electronic correlations remains secondary. Identification of the electronic-correlation footprints in topological phase diagrams of RSOC-coupled superconductors with a sizable electron-electron interaction remains thus an outstanding task.  

We carry out such an analysis within the framework of microscopic $t$-$J$-$U$ model \cite{Spalek2017}, supplemented with both RSOC and the Zeeman applied magnetic field. The $t$-$J$-$U$ Hamiltonian incorporates on-site Coulomb repulsion ($U$) and antiferromagnetic exchange interaction ($J$) simultaneously, generalizing the canonical Hubbard- and $t$-$J$-models and encompassing both of them as particular cases.  The model is analyzed within the framework of statistically-consistent Gutzwiller approximation (SGA), previously applied to inversion-symmetric correlated models of high-$T_c$ copper oxide superconductors \cite{Jedrak2010, Jedrak2011}. For fixed magnitude of kinetic exchange $J$, we construct the electronic density vs. $U$ phase diagram, for both hole- and electron doping and across the metal-to-insulator transition. It is shown that application of Zeeman field induces $C = - 4$ TSC close to the half-filling. Even though this topological state persists in the strong-coupling regime, we argue that electronic correlations are not instrumental for its stabilization. Our analysis of the $t$-$J$-$U$ model topological phase diagram reveals also a distinct region with $C = - 2$ TSC reached through the appearing concomitantly first-order metamagnetic and SC transitions. In this case, the change of topological invariant from $C = 0$ to $C = - 2$ is not accompanied by a closure of the bulk gap in the single-particle spectrum, highlighting the key role of electronic correlations. Such unconventional topological transformations taking place without bulk charge-gap closure have recently attracted attention in the context of MoTe$_2$/WSe$_2$ moiré heterobilayers.\cite{Li2021, Mai2024} Our work provides a concrete mechanism resulting in such behavior and based on a correlation-assisted first-order transition.

We also carry out a quantitative analysis of the singlet-triplet mixing as a function of electronic density (band filling), $n \equiv 1 - \delta$, with $\delta$ being the doping. Above the metal-to-insulator transition, the $p$-wave and $d$-wave SC amplitudes ($\Delta_p$ and $\Delta_d$, respectively) exhibit qualitatively distinct scaling, both with and without applied magnetic field. We obtain $\Delta_p \propto \delta^2$ and $\Delta_d \propto |\delta|$ close to half-filling, which indicates a strong suppression of the RSOC-induced triplet SC component on the underdoped side of high-$T_c$ phase diagram. 

The structure of the paper is as follows. In Sec.~\ref{section:model} we summarize main ingredients associated with the application of SGA method used here. In Sec.~\ref{section:order_parameter} we characterize types of order we discuss in the remaining part of the paper. Sec.~\ref{section:results} presents the results and their novel feature, whereas Sec.~\ref{section:summary} contains conclusions and outlook.

\section{Model and method}
\label{section:model}

We consider the Hamiltonian

\begin{align}
\label{eq:Hfinal}
 \hat{H} = \hat{H}_\textit{t-J-U} + \hat{H}_{\text{SO}} + \hat{H}_{\text{Zeeman}},  
\end{align}
\noindent
where

\begin{align}
\label{eq:HtJU}
 \hat{H}_\textit{t-J-U} &= {\sum_{ij\sigma}}^\prime t_{ij} \hat{a}_{i\sigma}^{\dagger} \hat{a}_{j\sigma} + U \sum_{i} \hat{n}_{i\uparrow} \hat{n}_{i\downarrow} \nonumber \\
  &+ J \sum_{\langle ij\rangle} \hat{\mathbf{S}}_{i} \cdot \hat{\mathbf{S}}_{j} - \mu \sum_{i\sigma} \hat{n}_{i\sigma}
\end{align}

\noindent
defines the $t$-$J$-$U$ model \cite{Spalek2022}. In Eq.~(\ref{eq:HtJU}) indices $i$ and $j$ enumerate sites of the square lattice, $\hat{a}_{i\sigma}$ ($\hat{a}_{i\sigma}^\dagger$) denote spin-$\sigma$ annihilation (creation) operators, and $\hat{n}_{i\sigma} \equiv \hat{a}_{i\sigma}^\dagger \hat{a}_{i\sigma}$. Spin operators $\hat{\mathbf{S}}_i \equiv (\hat{S}^x_i, \hat{S}^y_i, \hat{S}^z_i)$, with $\hat{S}_i^\alpha \equiv \frac{1}{2} \sum_{\sigma \sigma^\prime} \hat{a}_{i\sigma}^\dagger {\sigma}^\alpha_{\sigma\sigma^\prime} \hat{a}_{i\sigma^\prime}$ are defined in terms of Pauli matrices $\sigma^\alpha$. Here $\langle ij \rangle$ indicates summation over nearest-neighbor sites (witch each pair of indices counted only once), and prime-marked lattice summation is restricted to $i \neq j$. The $t$-$J$-$U$ model is governed by hopping integrals, $t_{ij}$, on-site Coulomb repulsion, $U$, as well as explicit antiferromagnetic exchange coupling, $J$ (independent of $U$). For brevity of notation, we have also incorporated the chemical potential term $\propto \mu$ directly into Eq.~(\ref{eq:HtJU}). The second term of Eq.~(\ref{eq:Hfinal}) represents the RSOC Hamiltonian

\begin{align}
\label{eq:Hso1}
 \hat{H}_{\text{SO}} = -V \sum_{i\sigma\sigma^{\prime}} [i(\hat{a}_{i\sigma}^{\dagger}\sigma_{\sigma\sigma^{\prime}}^{x}\hat{a}_{i+\hat{y}\sigma^{\prime}}-\hat{a}_{i\sigma}^{\dagger}\sigma_{\sigma\sigma^{\prime}}^{y}\hat{a}_{i+\hat{x}\sigma^{\prime}})+\mathrm{H.c.}],
\end{align}

\noindent
controlled by a single parameter $V$. We set lattice spacing to unity so that $\hat{x} \equiv (1, 0)$ and $\hat{y} = (0, 1)$ connect corresponding nearest-neighbor pairs. The term~(\ref{eq:Hso1}) is the only one in the total Hamiltonian~(\ref{eq:Hfinal}) that breaks the inversion symmetry. This becomes apparent by noting that transformation $\hat{x} \rightarrow - \hat{x}$ and $\hat{y} \rightarrow - \hat{y}$ implies $\hat{H}_{\text{SO}} \rightarrow - \hat{H}_{\text{SO}}$. Finally, the Zeeman term reads $\hat{H}_{\text{Zeeman}} = -h \sum_{i} (\hat{n}_{i\uparrow} - \hat{n}_{i\downarrow})$, with $h$ being the applied field.

Hereafter, we retain only hopping integrals between nearest- and next-nearest neighbors, $t \equiv -0.35\,\mathrm{eV}$ and $t^\prime \equiv 0.25 |t|$, respectively. The antiferromagnetic exchange is set to a constant value $J = 0.3 |t|$, whereas the on-site Coulomb repulsion is varied in the range $U = 0$-$14|t|$. This setup allows us to explore both the high-$T_c$ copper oxide regime ($U  \gtrsim W = 8 |t|$, with $W$ being the bare single-particle bandwidth), as well as  weak- and intermediate-coupling range ($U \lesssim W$). We consider RSOC magnitude $V = 0$-$0.1 |t|$, and Zeeman field $h = 0$-$0.02 |t|$. All simulations have been carried out at low temperature, $k_B T = 10^{-5} |t|$, with $k_B$ being Boltzmann constant.

The choice of Hamiltonian~(\ref{eq:HtJU}) as the basis for the present analysis results in several methodological advantages over commonly adopted Hubbard and $t$-$J$ models, both of which are encompassed as particular limits ($U \rightarrow \infty$ and $J \rightarrow 0$, respectively). Unlike the $t$-$J$ model, the $t$-$J$-$U$ Hamiltonian~(\ref{eq:HtJU}) does not involve cumbersome projection eliminating doubly-occupied site configurations, which is naturally attained in the limit $U \rightarrow \infty$. Moreover, Eq.~(\ref{eq:HtJU}) defines a tunable effective theory of formally simple structure, in contrast to the $t$-$J$ Hamiltonian derived from the microscopic Hubbard-type models via a proper canonical transformation \cite{Chao1977}. The $t$-$J$ model, in addition to the leading-order antiferromagnetic exchange, also incorporates other contributions, such as correlated hopping and ring exchange \cite{Delannoy2009}, with substantial numerical prefactors. In the presence of RSOC, the $t$-$J$ model needs to be supplemented also with Dzyaloshinskii-Moriya (DM) interactions \cite{Coffey1991, Nally2024}. While these terms are often neglected, this simplification turns the $t$-$J$ model into an effective theory with no direct advantages over the more general $t$-$J$-$U$ model. On the other hand, whereas the Hubbard model is free of those issues, its analysis within the variational scheme is more involved and requires going beyond the renormalized mean-field theory via either Monte-Carlo simulations or specialized real-space diagrammatic expansions \cite{Kaczmarczyk2014}. 
 
The model~\eqref{eq:Hfinal} is analyzed with the framework of the Statistically-Consistent Gutzwiller approximation (SGA) \cite{Spalek2022}. In the zero-temperature limit, SGA is based on minimization of the energy functional 

\begin{equation}
\label{eq:E_G}
    E_G \equiv \frac{\langle\Psi_G|\hat{H}|\Psi_G\rangle}{\langle\Psi_G|\Psi_G\rangle},
\end{equation}

\noindent
with respect to the variational wave function $|\Psi_G\rangle$ for fixed electron density. We adopt $|\Psi_G\rangle = \hat{P}_G |\Psi_0\rangle$ expressed in terms of the Slater determinant state $|\Psi_0\rangle$ (possibly representing the broken-symmetry SC and magnetic solution) and the correlator operator $\hat{P}_G \equiv \prod_{i} \hat{P}_{Gi} |\Psi_0\rangle$ in the product form. The operators

\begin{align}
\label{eq:projector_gen}
\hat{P}_{Gi} \equiv \lambda^0_i|0\rangle_{ii}\langle0|+\sum_{\sigma\sigma^\prime}\lambda^{\sigma\sigma^\prime}_i|\sigma\rangle _{ii}\langle{\sigma}|+\lambda^d_i|d\rangle_{ii}\langle d|
\end{align}

\noindent
are introduced to readjust the weights of local configurations in response to interactions. In Eq.~\eqref{eq:projector_gen} $|0\rangle$, $|{\uparrow}\rangle_i$, $|{\downarrow}\rangle_i$, and $|d\rangle_i \equiv |{\uparrow\downarrow}\rangle_i$ comprise the local basis states at the lattice site $i$, composed of empty ($|0\rangle$), singly and doubly occupied ($|{\sigma}\rangle_i$) and doubly occupied ($|{d}\rangle_i$) states. The parameters $\lambda^0_i$, $\lambda^{\sigma\sigma^\prime}_i$, and $\lambda^d_i$ are further constrained by the conditions $\langle \Psi_0| \hat{P}_{Gi}^2 | \Psi_0\rangle = 1$ and $\langle \Psi_0| \hat{P}_{Gi} \hat{a}^\dagger_{i\sigma} \hat{a}_{i\sigma^\prime} \hat{P}_{Gi} | \Psi_0\rangle = \langle \Psi_0| \hat{a}^\dagger_{i\sigma} \hat{a}_{i\sigma^\prime} | \Psi_0\rangle$, which allows for efficient Wick's decomposition of the $E_G$ functional \cite{Buenemann2012,Fidrysiak2021}. Here we retain only leading-order diagrams in the Wick's expansion, which results in the SGA approximation. Generalization to finite temperatures amounts to consideration of the free energy functional in place of $E_G$ \cite{Jedrak2010}. We note that one can often disregard the off-diagonal correlator component $\propto \lambda^{\uparrow\downarrow}_i |{\uparrow}\rangle_{ii} \langle{\downarrow}| + \lambda^{\downarrow\uparrow}_i |{\downarrow}\rangle_{ii} \langle{\uparrow}|$ in Eq.~\eqref{eq:projector_gen}, particularly in the paramgnetic and collinear magnetic phases. However, in the presence of spin-orbit coupling those terms cannot be outright eliminated due to mixing between spin-up and spin-down states. Yet, RSOC term~(\ref{eq:Hso1}) is antisymmetric in wave-vector space and we consider specifically the Zeeman field applied along $z$-axis. In turn, while spin off-diagonal hopping amplitudes generally emerge on non-local bonds, the structure of the local spin-nondiagonal expectation values remain zero, i.e., $\langle \hat{a}_{i\uparrow}^\dagger \hat{a}_{i\downarrow} \rangle = 0$. Those properties of local density matrix are reflected also in optimized local correlator as $\lambda^{\uparrow\downarrow}_i = \lambda^{\downarrow\uparrow}_i = 0$. When analyzing the general correlator of Eq.~(\ref{eq:projector_gen}), we have verified that the latter condition is satisfied in our numerical simulations.

In the present study we restrict the variational space to states compatible with the lattice translational symmetry, thereby excluding antiferromagnetic, spin-density-wave (SDW) and charge-density-wave (CDW) orderings. The magnetic and SC order parameters are otherwise not restricted. This provides a minimal setup for the discussion of TSC in applied Zeeman field, and admits calculations for lattice size large enough to account for SC gaps on sub-meV scale. The structure of the resultant SC order parameter is detailed in Sec.~\ref{section:order_parameter} below, whereas the discussion of plausible extensions, involving intertwined SDW/CDW and SC orders, are postponed to Sec.~\ref{section:summary}.

\section{Superconducting order parameter}
\label{section:order_parameter}
We construct the dominant $d_{x^2-y^2}$-wave SC order parameter as

\begin{align}
\label{eq:Delta_d}
 \Delta_{d} = \frac{1}{4} \left[ (F^{\downarrow\uparrow}_{+\hat{x}} + F^{\downarrow\uparrow}_{-\hat{x}}) - (F^{\downarrow\uparrow}_{+\hat{y}} + F^{\downarrow\uparrow}_{-\hat{y}}) \right],
\end{align}
where $F_{\delta}^{\sigma\sigma^\prime} \equiv \frac{1}{N} \sum_i \langle \hat{a}_{i+\delta, \sigma}^{\dagger} \hat{a}^{\dagger}_{i, \sigma^\prime} \rangle$ denotes the bond pairing amplitude between neighboring lattice sites ($\delta = \pm \hat{x}, \pm \hat{y}$), and $N$ is the lattice size. We have introduced a multiplicative factor of $\frac{1}{4}$ so that the magnitude of $\Delta_{d}$ is the same as that of bond amplitudes. The relative minus sign between the spatial components $F^{\downarrow\uparrow}_{\pm\hat{x}}$ and $F^{\downarrow\uparrow}_{\pm\hat{y}}$ reflects the $d_{x^2 - y^2}$ symmetry. 

\begin{figure}
    \centering
    \includegraphics[width=1.0\linewidth]{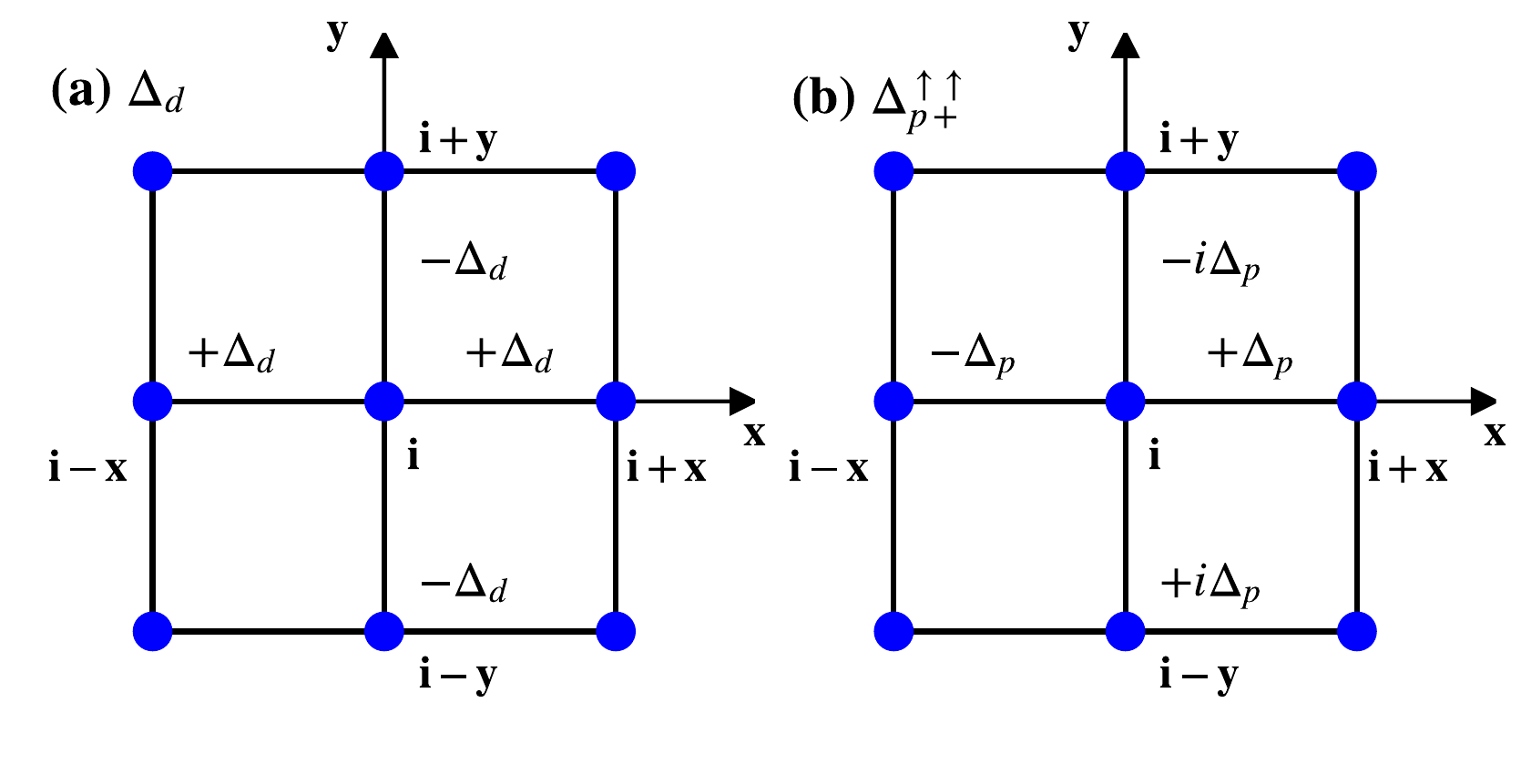}
    \caption{Schematic real-space representation of SC bond amplitudes for (a) singlet and (b) one of triplet components of the mixed parity order parameter, $\Delta_d$ and $\Delta^{\uparrow\uparrow}_{p_+}$, respectively. Lattice sites are marked by blue circles. The singlet amplitude remains invariant under spatial inversion and acquires phase $\pi$ (equivalent to sign reversal) under spatial rotation by $\theta = \pi/2$. (b) Equal-spin amplitudes contributing to $\Delta^{\uparrow\uparrow}_{p_+}$ order parameter component are odd under inversion and acquire a relative phase of $-\pi/2$ under $\theta = \pi/2$ spatial rotation. The remaining nonzero amplitude, $\Delta^{\downarrow\downarrow}_{p_-}$, is not shown as it exhibits phase structure analogous to that illustrated in (b), yet acquires opposite phase upon circulation of the central site, $i$.}
    \label{fig:phase_order}
\end{figure}

The presence of RSOC in Eq.~(\ref{eq:Hfinal}) breaks spatial inversion symmetry, thereby parity-mixing the pairing channels and inducing a triplet component. Subsequently, chiral equal-spin $p$-wave order parameters are constructed as

\begin{align} \label{eq:Delta_p}
    \Delta_{p_\pm}^{\sigma\sigma} =\frac{1}{4}  \left[F_{+\hat{x}}^{\sigma\sigma} - F_{-\hat{x}}^{\sigma\sigma} \pm i\,( F_{+\hat{y}}^{\sigma\sigma} - F_{-\hat{y}}^{\sigma\sigma})\right].
\end{align}

\noindent
The amplitudes in Eq.~\eqref{eq:Delta_p} reflect the formation of Cooper pairs with total $z$-axis spin projection $S^z = \pm 1$. We note that $S^z = 0$ triplet amplitude is also admitted in a general scenario, particularly for Zeeman field tilted away from the $z$-axis. We have, however, verified that this component remains zero for the setup considered in the present study. The phase structure of singlet- and triplet pairing amplitudes comprising the singlet- and triplet- components of the order parameter in real space is illustrated in Fig.~\ref{fig:phase_order}.

\section{Results}
\label{section:results}

\subsection{Superconductivity without Zeeman field}
\label{subsection:Superconductivity without Zeeman field}

\begin{figure}
    \centering
    \includegraphics[width=1.0\linewidth]{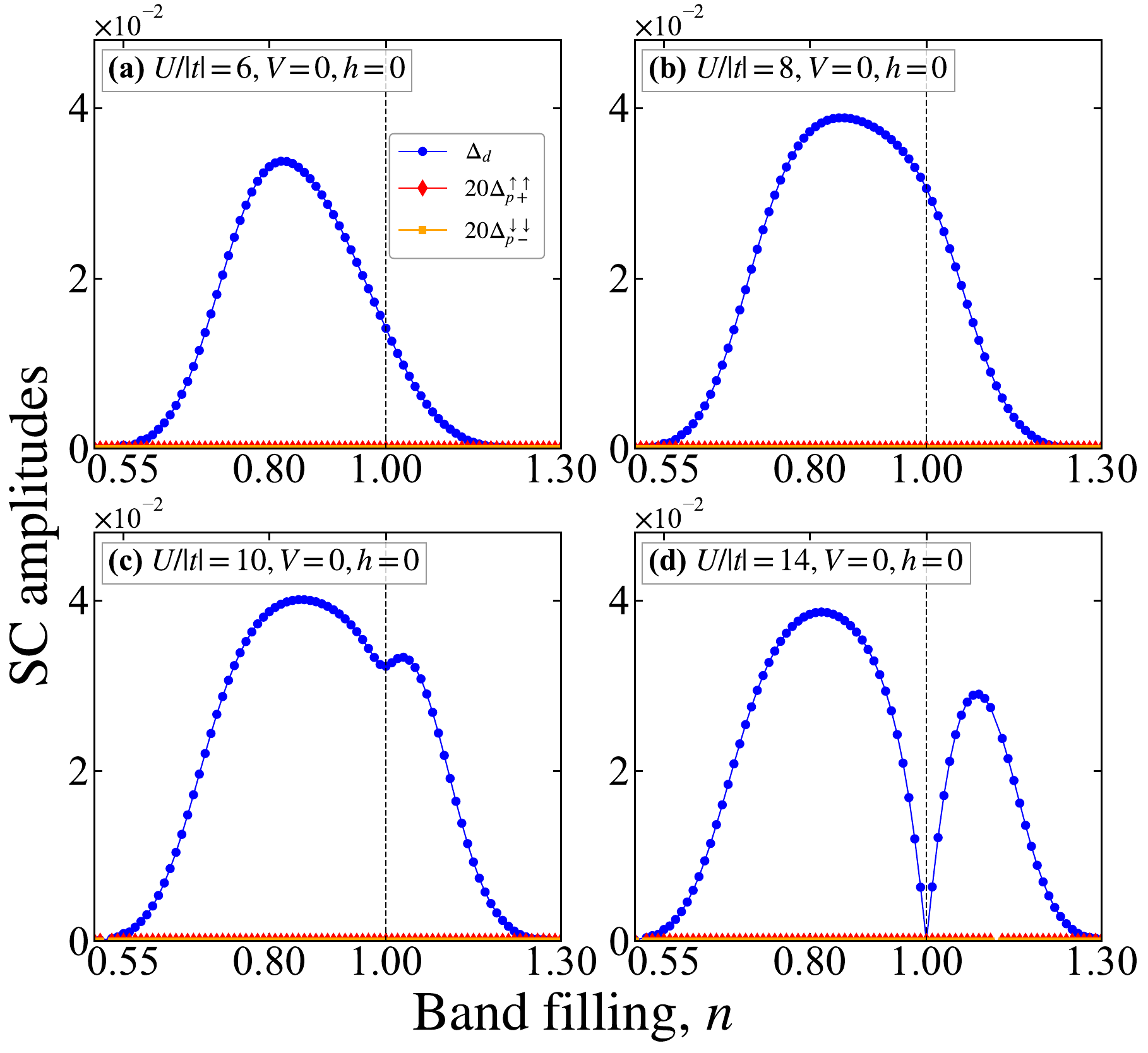}
    \caption{Reference results for spin-orbit coupling and Zeeman field set to zero ($V \equiv 0$, $h \equiv 0$). The panels show the band-filling dependence of the singlet $d$-wave order parameter, $\Delta_d$, as well as equal-spin triplet $p$-wave components, $\Delta_{p+}^{\uparrow\uparrow}$ and $\Delta_{p-}^{\downarrow\downarrow}$ (here equal to zero, indicating pure $d$-wave SC). The on-site Coulomb interaction is set to (a) $U=6 |t|$, (b) $U=8 |t|$, (c) $U=10 |t|$, and (d) $U=14 |t|$. The remaining parameters are $t = -0.35\,\mathrm{eV}$, $t^\prime = 0.25 |t|$, $J = 0.3 |t|$, $k_B T = 10^{-5} |t|$, and lattice of size has been set to $N=800 \times 800$. Metal-to-insulator transition at the half filling ($n = 1$) occurs in between $U = 10 |t|$ and $U = 14 |t|$, cf. panels (c) and (d).}
    \label{fig:SC_amp_V0.0_h0.0}
\end{figure}

We first address the reference situation in the absence of applied Zeeman field ($h = 0$) and RSOC ($V = 0$). In Fig.~\ref{fig:SC_amp_V0.0_h0.0} we show the calculated $d$-wave SC order parameter ($\Delta_d$) and relevant triplet components ($\Delta_{p+}^{\uparrow\uparrow}$, $\Delta_{p-}^{\downarrow\downarrow}$) as a function of band filling. The panels correspond to the on-site Coulomb repulsion (a) $U=6 |t|$, (b) $U=8 |t|$, (c) $U=10 |t|$, and (d) $U=14 |t|$. The remaining parameters are kept fixed to $t = - 0.35\,\mathrm{eV}$, $t^\prime = 0.25 |t|$, $J = 0.3 |t|$, $k_B T = 10^{-5} |t|$, and the calculations have been carried out for the lattice size $N=800 \times 800$. In the absence of RSOC, SC is of definite (even) parity and all triplet SC amplitudes vanish identically. Increase of the on-site Coulomb repulsion $U$ leads to gradual formation of two SC domes: one on the hole- and another on the electron-doped side of the phase diagram. At half-filling ($n=1$), strong correlations trigger a Brinkman-Rice metal-to-insulator transition at $U \gtrsim 12 |t|$ \cite{Brinkman1970}. The latter manifests itself as a suppression of all SC amplitudes at $n = 1$, cf. Fig.~\ref{fig:SC_amp_V0.0_h0.0}(c) and (d), whereas for $n\neq 1$ SC emerges on both hole end electron sides of the phase diagram. The $t$-$J$-$U$ model capability to account for SC evolution across metal-to-insulator transition illustrates the advantage of simultaneous incorporation of both $U$- and $J$-terms in Eq.~(\ref{eq:HtJU}) \cite{Jedrak2011}.

\begin{figure}
    \centering
    \includegraphics[width=1\linewidth]{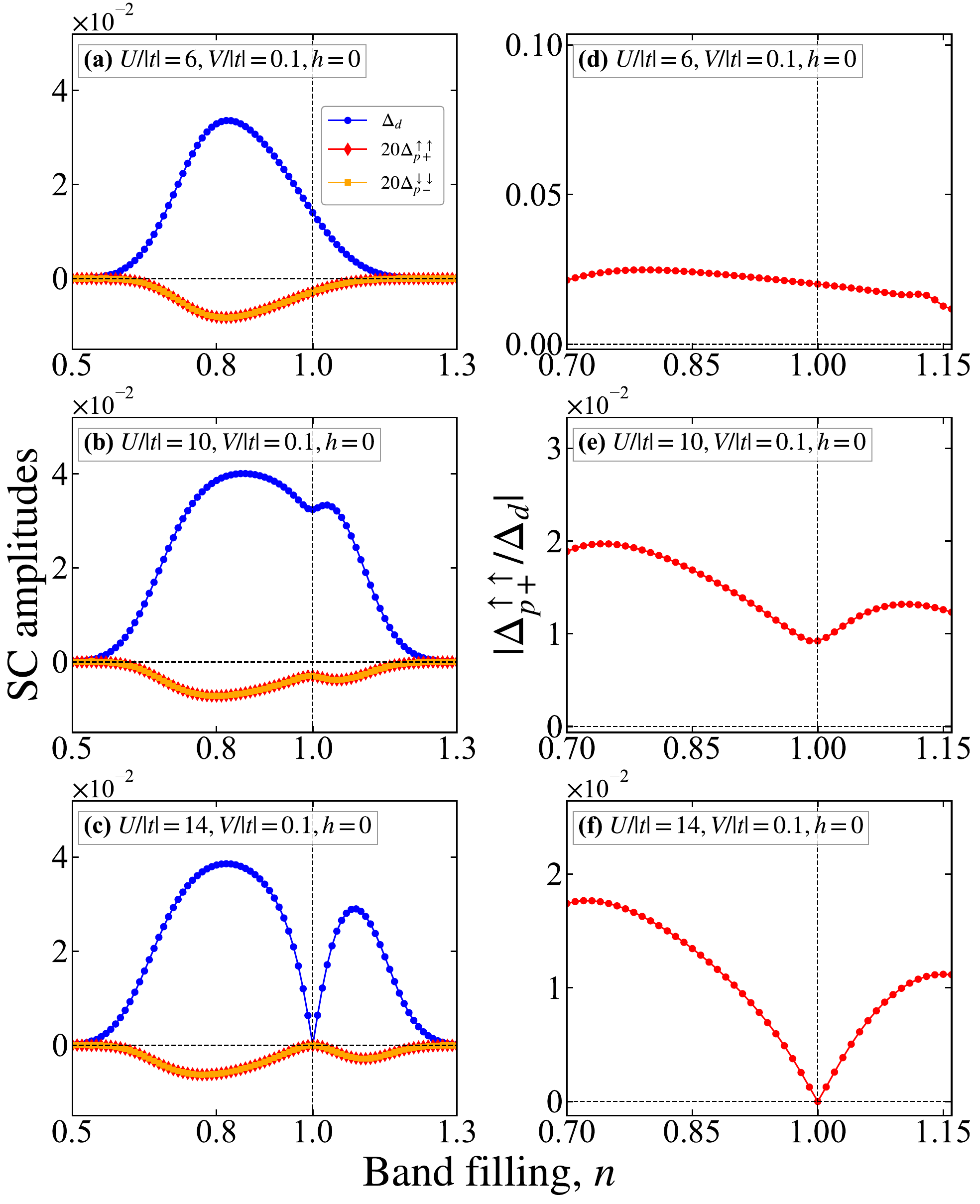}
    \caption{Characteristics of the mixed-parity SC as a function of electronic density for nonzero RSOC magnitude ($V = 0.1 |t|$) and no applied Zeeman field ($h = 0$). The on-site Coulomb repulsion is set to $U=6 |t|$ [(a) and (d)], $U=10 |t|$ [(b) and (e)], and $U=14 |t|$ [(c) and (f)], whereas the remaining parameters are the same as those adopted in Fig.~\ref{fig:SC_amp_V0.0_h0.0}. Left panels detail singlet $d$-wave order parameter, $\Delta_d$, alongside the relevant equal-spin triplet $p$-wave components, $\Delta_{p+}^{\uparrow\uparrow}$ and $\Delta_{p-}^{\downarrow\downarrow}$ (multiplied by a factor of 20 for clarity). Equal magnitudes of $\Delta_{p+}^{\uparrow\uparrow}$ and $\Delta_{p-}^{\downarrow\downarrow}$ amplitudes are a consequence of time-reversal symmetry. Right panels show the corresponding ratios $|\Delta^{\uparrow\uparrow}_{p_+}/\Delta_d|$ (equivalent to $|\Delta^{\downarrow\downarrow}_{p_-}/\Delta_d|$). Panel (f) indicates a qualitatively distinct scaling of singlet- and triplet SC components close to half-filling in the strong-correlation regime. The SC amplitudes are composed of two-point anomalous expectation values and are thus dimensionless.}
    \label{fig:SC_amp_ratio_V0.1_h0.0}
\end{figure}

We now consider the case of nonzero RSOC of magnitude $V=0.1 |t| = 35\,\mathrm{meV}$, within the range of empirically estimated $V \gtrsim 10\,\mathrm{meV}$ for representatives of Bi-family of high-$T_c$ cuprates~\cite{Gotlieb2018, Luo2024}.  Figure~\ref{fig:SC_amp_ratio_V0.1_h0.0}(a)-(c) details the band-filling dependence of the calculated $d$-wave and $p$-wave SC amplitudes for increasing values of the on-site Coulomb repulsion, $U$, across the metal-to-insulator transition. The adopted values of $U$ are detailed inside the panels, and the $p$-wave amplitude has been multipled by a factor of $20$ for clarity. The explicit breakdown of spatial inversion symmetry by RSOC induces parity mixing, generating finite equal-spin chiral $p$-wave amplitudes along with the dominant $d$-wave singlet channel. We note that, owing to the spin-momentum locking dictated by the RSOC, only specific symmetry-matched pairs out of the four possible chirality-spin configurations prevail. The $|{\uparrow\uparrow}\rangle$ spin state couples exclusively to the positive chirality channel ($\Delta_{p_+}^{\uparrow\uparrow}$), while the $|{\downarrow\downarrow}\rangle$ state couples to the negative chirality channel ($\Delta_{p_-}^{\downarrow\downarrow}$), as formally expressed in Sec.~\ref{section:order_parameter}. The remaining components, $\Delta_{p_-}^{\uparrow\uparrow}$ and $\Delta_{p_+}^{\downarrow\downarrow}$, are identically zero. Moreover, in the absence of Zeeman field time-reversal symmetry is preserved so that the nonzero chiral triplet components remain equal, $\Delta_{p_+}^{\uparrow\uparrow} = \Delta_{p_-}^{\downarrow\downarrow}$. As is apparent in Fig.~\ref{fig:SC_amp_ratio_V0.1_h0.0}, direct simulation satisfies these conditions, which also validates our variational scheme. Apart from the onset of the parity-mixed state, the phase diagram remains qualitatively consistent with that obtained for $V=0$ (cf. Fig.~\ref{fig:SC_amp_V0.0_h0.0}).

As seen in Fig.~\ref{fig:SC_amp_ratio_V0.1_h0.0}(c), with increasing the Hubbard $U$ the system evolves continuously through the metal-to-insulator-transition region. However, singlet and triplet SC amplitudes exhibit a qualitatively distinct behavior close to half-filling. This is illustrated in Fig.~\ref{fig:SC_amp_ratio_V0.1_h0.0}(d)-(f), detailing of the ratio $|\Delta^{\uparrow\uparrow}_{p_+} / \Delta_d|$ as a function of electronic density $n$ for the values of $U$ corresponding to those adopted in panels (a)-(c). Note that $|\Delta^{\downarrow\downarrow}_{p_-}| = |\Delta^{\uparrow\uparrow}_{p_+}|$ for $h = 0$ so this quantity is the same for both triplet components. As $n \rightarrow 1$, $|\Delta^{\uparrow\uparrow}_{p_+} / \Delta_d|$ approaches a constant value below the Brinkman-Rice transition [panels (d) and (e)], whereas $|\Delta^{\uparrow\uparrow}_{p_+} / \Delta_d| \rightarrow 0$ above that point [panel (f)]. This result provides insight into the odds of observing the mixed parity SC footprints across phase diagrams of materials characterized by varying interaction strength. Specifically, it implies that detection of the triplet SC component in strongly-correlated superconductors might be challenging in the underdoped regime, where triplet SC amplitudes are strongly suppressed. Indeed, above the Brinkman-Rice transition point we find that $|\Delta_d| \propto |\delta| $ and $|\Delta_{p_+}| \propto \delta^2$, where $\delta \equiv 1-n$ is the doping level. This is consistent with $|\Delta_{p_+}/\Delta_d| \propto |\delta|$ scaling close to half-filling seen in Fig.~\ref{fig:SC_amp_ratio_V0.1_h0.0}(c). The situation is qualitatively different for weakly and intermediately correlated systems, where the triplet component is approximately proportional to its singlet counterpart across entire phase diagram. 

\begin{figure}
    \centering
    \includegraphics[width=1\linewidth]{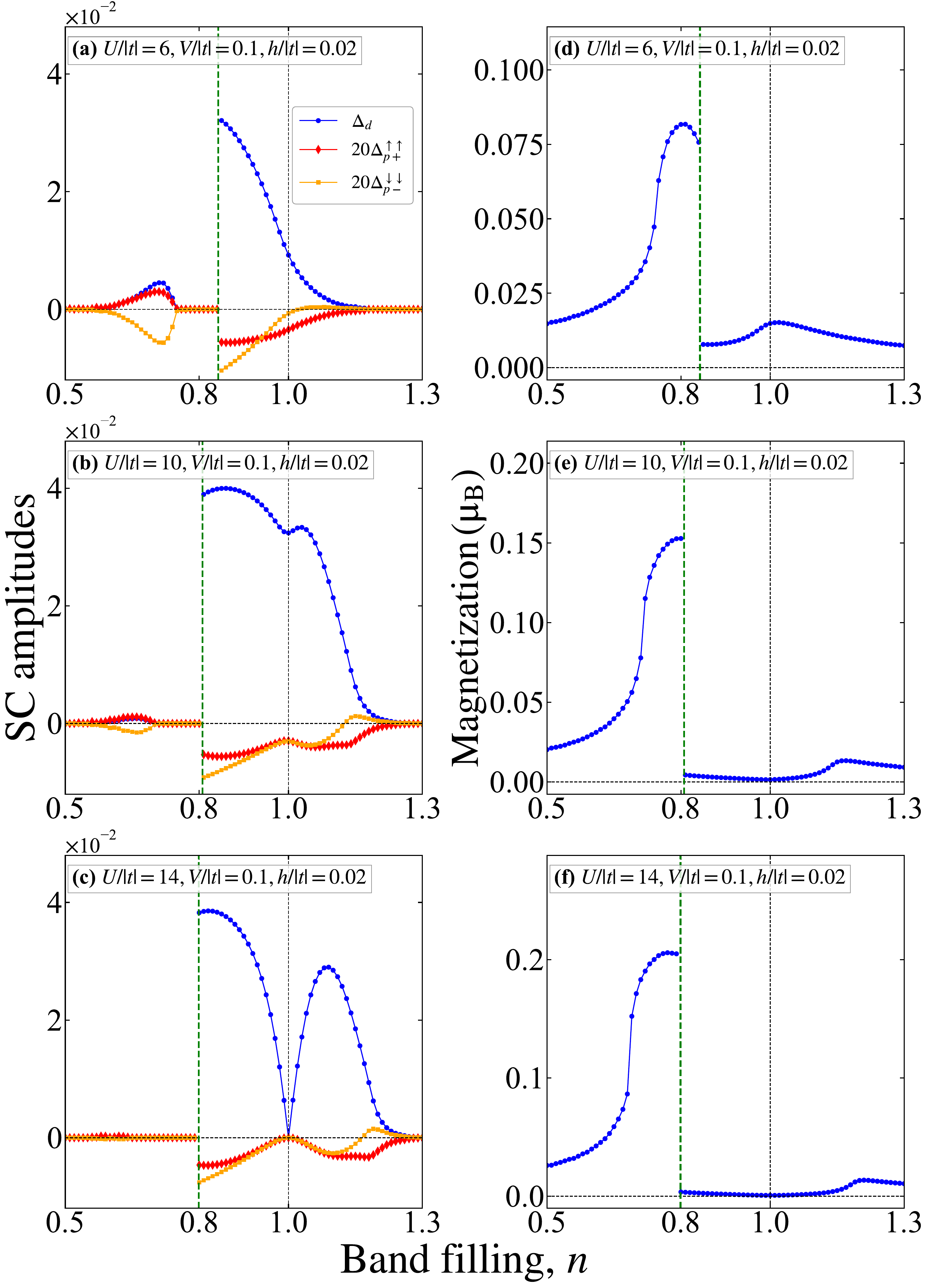}
    \caption{Band-filling dependence of the SC and magnetic properties for nonzero RSOC and Zeeman field ($V = 0.1$, $h=0.02$). The panels are organized according to increasing on-site Coulomb repulsion: $U=6 |t|$ [(a), (d)], $U=10 |t|$ [(b), (e)], and $U=14 |t|$ [(c), (f)]. The remaining parameters are the same as those adopted in Sec.~\ref{subsection:Superconductivity without Zeeman field}. Left panels detail the band-filling evolution of the singlet ($\Delta_d$) and triplet ($\Delta_{p+}^{\uparrow\uparrow}$, $\Delta_{p-}^{\downarrow\downarrow}$) SC amplitudes, whereas right panels show the magnetization. A discontinuity in both the SC amplitudes and the magnetization near $n \approx 0.8$ for all considered values of $U$ indicates first-order phase transition (marked by vertical dashed lines).}
    \label{fig:SC_and_Mag_V0.1_h0.02}
\end{figure}

\subsection{Superconductivity in Zeeman field}
\label{subsection:Superconductivity with Zeeman field}

\begin{figure*}
  \centering
  \includegraphics[width=1\linewidth]{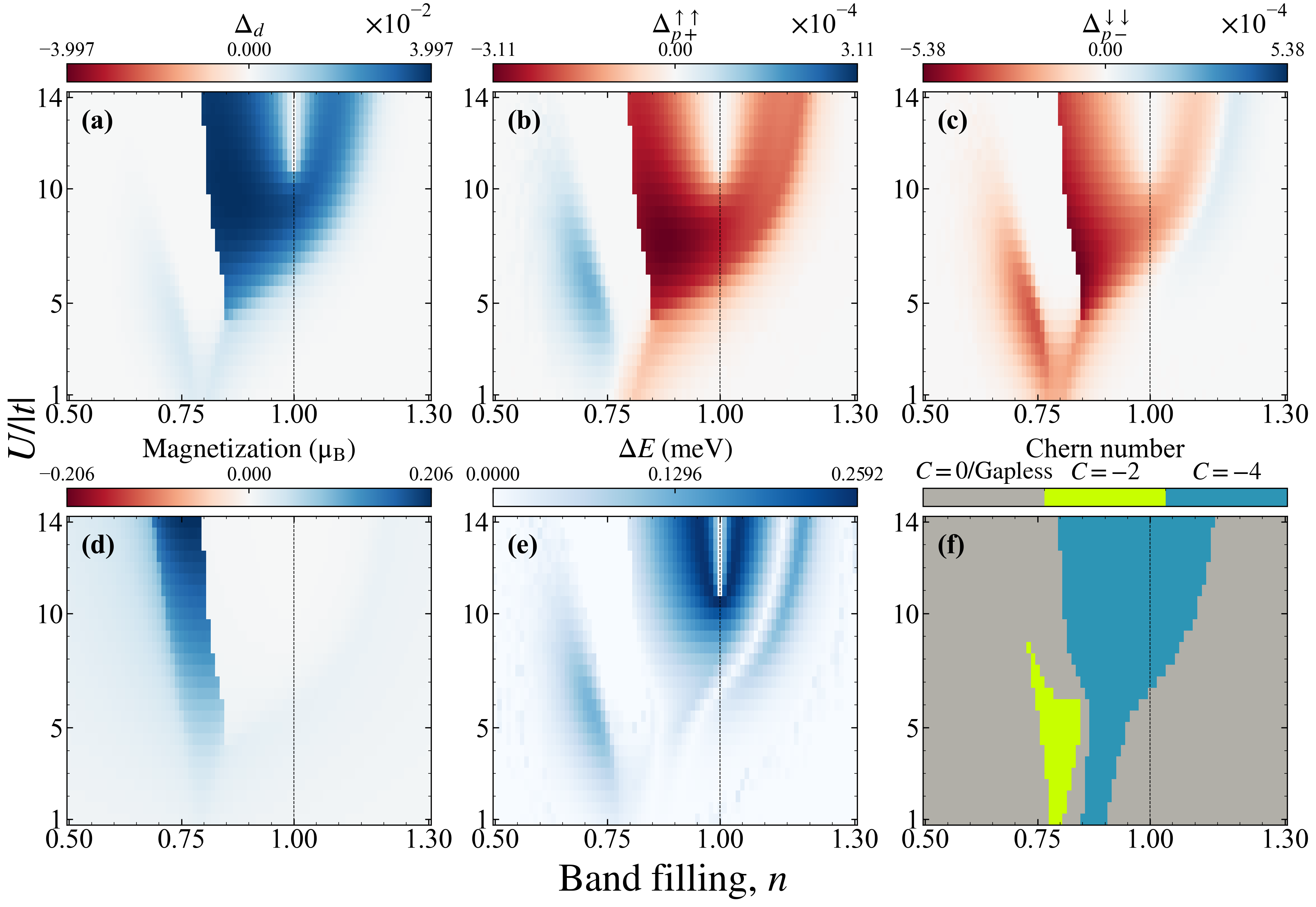}
  \caption{Band filling vs. on-site Coulomb repulsion phase diagram of RSOC-coupled $t$-$J$-$U$ model in an applied Zeeman field. The adopted parameters are $t = -0.35\,\mathrm{eV}$, $t^\prime = 0.25 |t|$, $J = 0.3 |t|$, $V=0.1$, $h=0.02$, $k_B T = 10^{-5} |t|$, and lattice size is set to $N = 800 \times 800$. The top row shows (a) the singlet $d$-wave amplitude $\Delta_d$, as well as equal-spin triplet components (b) $\Delta_{p_+}^{\uparrow\uparrow}$ and (c) $\Delta_{p_-}^{\downarrow\downarrow}$. The bottom row details the corresponding (d) magnetization (in $\mu_B$), (e) bulk energy gap $\Delta E$ (in meV), and (f) Chern number $C$. A sharp discontinuity is apparent across the parameter space in panels (a-e) for $n \approx 0.8$, signaling a first-order phase transition. In particular, the $C=-2$ state is reachable via the first-order transition without a single-particle gap closure. The  $C=-4$ region remains robust close to half-filling in a broad range of $U$. }
  \label{fig:Phase_D_V0.1_h0.02}
\end{figure*}

We now turn to the discussion of the joint effects of RSOC and electronic correlations for $V = 0.1 |t|$ and $h = \mu_B B = 0.02 |t|$, incorporating all terms of the Hamiltonian~(\ref{eq:Hfinal}). The remaining parameters are the same as those adopted in Sec.~\ref{subsection:Superconductivity without Zeeman field}. In Fig.~\ref{fig:SC_and_Mag_V0.1_h0.02}, we present calculated SC amplitudes (left panels) and magnetization (right panels) for on-site Coulomb repulsion $U=6 |t|$ [(a), (d)], $U=10 |t|$ [(b), (e)], and $U=14 |t|$ [(c), (f)]. Since the Zeeman field breaks time-reversal symmetry explicitly, the degeneracy the chiral triplet SC components, $\Delta_{p+}^{\uparrow\uparrow}$ and $\Delta_{p-}^{\downarrow\downarrow}$, is lifted. A prominent feature seen in Fig.~\ref{fig:SC_and_Mag_V0.1_h0.02} is a discontinuity in both the SC amplitudes [panels (a)-(c)] and the corresponding magnetization [panels (d)-(f)] on the hole-doped side of the phase diagram for $n \approx 0.8$. This is indicative of a simultaneous first-order metamagnetic and SC phase transitions, marked by a vertical dashed line. The transition point has been determined based on minimization of the free energy functional, as explained in Appendix~\ref{appendix:hysteresis}, where we also demonstrate a broad phase coexistence regime around the transition. Moreover, Fig.~\ref{fig:SC_and_Mag_V0.1_h0.02} evidences a competition between SC and magnetism. Pairing amplitudes in the high-moment state ($n \lesssim 0.8$) are substantially suppressed with respect to those in the low-moment phase on the right-hand side of the transition. In brief, experimental observation of this SC-metamagnetism coexistence achieved via first-order transition would provide a strong evidence of the correlation- and spin-orbit-cooperative effects in such quasi-two-dimensional systems.

Extending the construction presented in Fig.~\ref{fig:SC_and_Mag_V0.1_h0.02}, we have composed a complete band filling vs. on-site Coulomb repulsion phase diagram in Fig.~\ref{fig:Phase_D_V0.1_h0.02}. The considered range of $U = 1$--$14 |t|$ represents the evolution from weak- to strong-correlation regime, that we have also verified explicitly by investigating the $U \rightarrow \infty$ ($t$-$J$-model) limit, cf. Subsection.~\ref {subsection: t-J model} below. The upper row [panels (a)-(c)] shows the evolution of SC amplitudes $\Delta_{d}$, $\Delta_{p_+}^{\uparrow\uparrow}$, and $\Delta_{p_-}^{\downarrow\downarrow}$. Blue- and red colors represent the relative phase difference between singlet- and triplet amplitudes (here either $0^\circ$ or $180^\circ$, which corresponds to $\pm 1$ relative sign. The latter cannot be removed by a global gauge transformation and is thus of physical significance. Both $p$-wave amplitudes exhibit sign reversals relative to the primary $d$-wave channel throughout the phase diagram. 

Figure~\ref{fig:Phase_D_V0.1_h0.02}(d) maps the magnetization induced by Zeeman field, which reaches a maximum of approximately $0.2\mu_B$ per site in the analyzed region. Figure~\ref{fig:Phase_D_V0.1_h0.02}(e) details the single-particle bulk energy gap, defined as a minimum energy difference between occupied and empty bands across the Brillouin zone, $\Delta E$. We emphasize that both the pure $d$-wave and mixed-parity SC states discussed in Sec.~\ref{subsection:Superconductivity without Zeeman field} are gapless in this sense, since time-reversal symmetry protects the nodal points in the quasiparticle spectrum. The nonzero gap seen in Fig.~\ref{fig:Phase_D_V0.1_h0.02}(e) is thus a consequence of the interplay between RSOC and Zeeman field. Unless the applied field is nonphysically large, the gap magnitude falls into the sub-meV range, which requires consideration of large lattices. This becomes apparent by referring to Anderson criterion \cite{Anderson1959}, relating mean spacing between energy levels to SC. Assuming the nodal gap $\Delta E \sim 10^{-4} |t| = 0.035\,\mathrm{meV}$ [cf. Fig.~\ref{fig:Phase_D_V0.1_h0.02}(e)], one can estimate the appropriate lattice size as $W / N \lesssim 10^{-4} |t|$ so that $N \gtrsim 283 \times 283$ (here $W = 8|t|$ is bare bandwidth scale). Even though adopted $N = 800 \times 800$ fulfills this condition with due margin, small yet noticable fluctuations due to finite size effects persist on the boudary of the SC state [cf. Fig.~\ref{fig:Phase_D_V0.1_h0.02}(e)]. A more detailed analysis of these aspects is presented in Appendix~\ref{appendix:finite_size}. The resultant fully gapped SC states are classified according to Chern number, $C$, which we carry out in Fig.~\ref{fig:Phase_D_V0.1_h0.02}(f). Two distinct TSC regions are identified, $C = -4$ TSC (blue) close to half-filling in broad range of on-site Coulomb repulsion, and $C = -2$ TSC (green) on hole-doped side of the phase diagram in weak- to moderate-coupling regime. Grey color marks both gapless state (numerically determined by the condition $\Delta E < 10^{-4}\,\mathrm{meV}$ and fully gapped, but topologically trivial ($C = 0$) SC. The Chern number has been calculated using an efficient Brillouin zone triangulation scheme, described in Ref.~\citenum{Fukui2005}. 

\begin{figure*}
 \centering
  \includegraphics[width=1\linewidth]{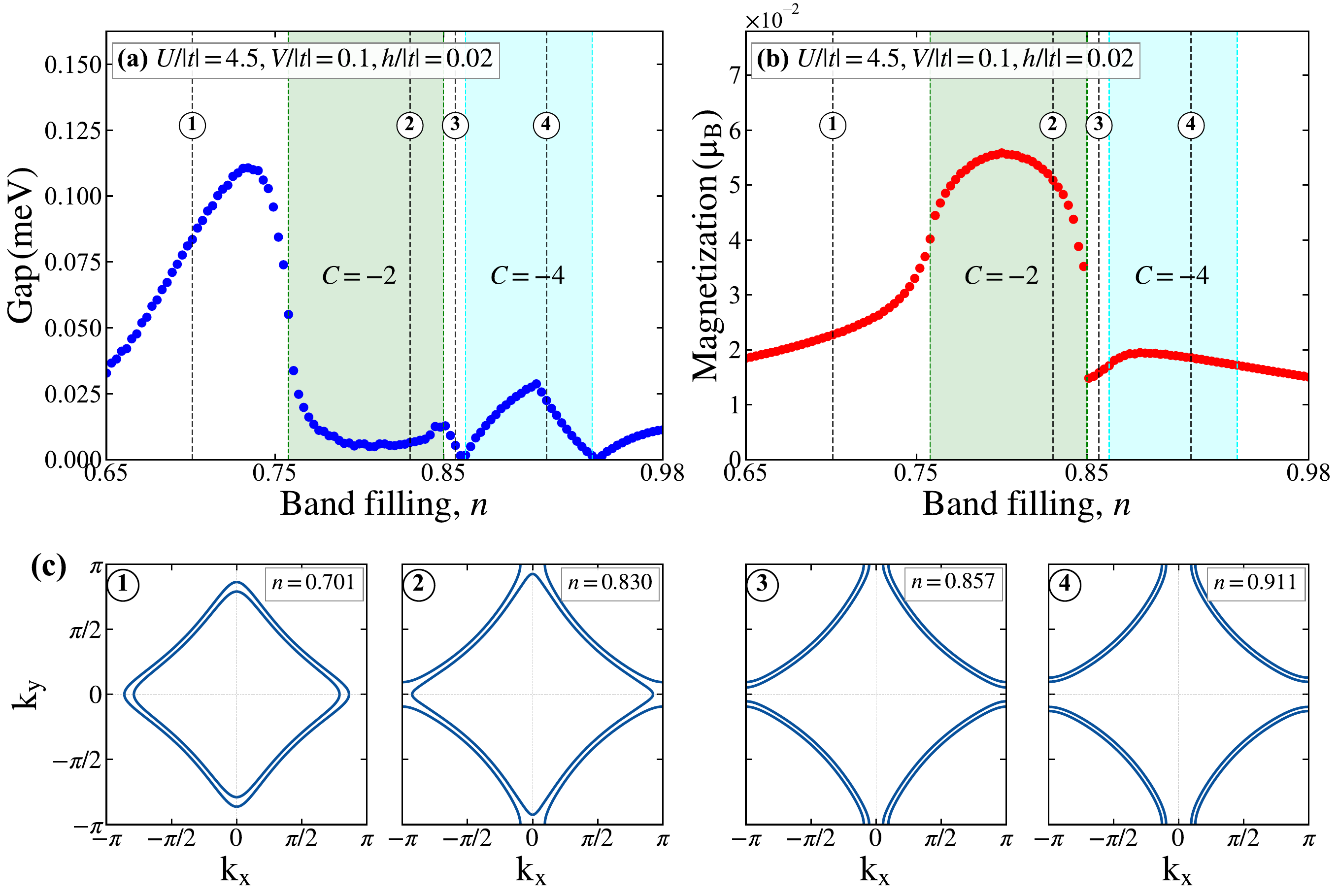}
  \caption{Band-filling evolution of SC, magnetic, and single-particle electronic properties for $U = 4.5 |t|$, corresponding to a horizontal cut within the phase diagram in Fig.~\ref{fig:Phase_D_V0.1_h0.02}. The remaining parameters are set to $t = -0.35$\,eV, $t^\prime = 0.25 |t|$, $J = 0.3 |t|$, $V=0.1|t|$, $h=0.02|t|$, $k_B T = 10^{-5} |t|$, and $N = 800 \times 800$. Top panels detail (a) the SC gap and (b) magnetization. Blue- and green-shaded regions mark the $C = -4$ and $C = -2$ TSC states, respectively. To correlate the topological transitions with the evolution of the single-particle electronic structure, panel (c) shows the underlying Fermi surface (as defined in the text) across the phase diagram. The selected densities, $n=0.701, 0.830, 0.857$, and $0.911$ (numbered as 1-4), are marked in panels (a) and (b) by dashed vertical lines for reference. Lifshitz transition associated with the emergence of $C = -2$ TSC is observed. }
  \label{fig:Gap_mag_Fermi}   
\end{figure*}

\subsection{Correlation-assisted topological transitions}

We now turn to the discussion of the electronic correlation effects on TSC, whose footprints may be identified in Fig.~\ref{fig:Phase_D_V0.1_h0.02}. Specifically, all the SC amplitudes [panels (a)-(c)] and magnetization [panel (d)] are discontinuous as a function of doping, with apparent jump at $n \approx 0.8$ for $U \gtrsim 4 |t|$ that is also associated with a change of Chern number [cf. panel (f)]. To scrutinize further this portion of the phase diagram and clarify the evolution of SC state topology, in Fig.~\ref{fig:Gap_mag_Fermi} we plot the band-filling dependence of SC gap [panel (a)] and magnetization [panel (b)] for fixed on-site Coulomb repulsion $U = 4.5|t|$, which amounts to a horizontal cut of Fig.~\ref{fig:Phase_D_V0.1_h0.02}(d) and (e). The considered range of electronic density encompasses both $C = -4$ and $C = -2$ TSC states marked in Fig.~\ref{fig:Phase_D_V0.1_h0.02}(f). Figure~\ref{fig:Gap_mag_Fermi}(a) shows that SC gap approaches zero at the boundaries of the $C = -4$ TSC (blue-shaded region), indicating that this state is reached via a conventional topological phase transition. Namely, the closure of the SC gap admits change of the Chern number that is otherwise robust against adiabatic perturbations of the Hamiltonian. Even though the $C = -4$ TSC region persists in a broad range of on-site Coulomb repulsion and is found also in the strong-coupling limit (cf. Sec.~\ref{subsection: t-J model} below), we thus conclude that electronic correlations are not instrumental for its appearance. Indeed, its presence has been reported also within the BdG scheme \cite{Yoshida2016,Daido2016}, which entirely disregards electron-electron interactions. The case of $C = -2$ TSC state [green region in Fig.~\ref{fig:Gap_mag_Fermi}(a)] is qualitatively different in this respect, as the single-particle gap does not close at its boundaries. Moreover, Fig.~\ref{fig:Gap_mag_Fermi}(b) shows that the magnetization jumps on the right-hand side of the $C=-2$ region ($n \approx 0.85$). We have confirmed, by carrying out forward and backward sweeps of the phase diagram, that this is a consequence of a correlation-driven first-order transition characterized by a broad region of the phase coexistence. The transition point has been determined by minimization of the free energy functional, as detailed in Appendix~\ref{appendix:hysteresis}. Remarkably, despite a reproducible change of Chern number on the left-hand side of the $C = -2$ TSC state ($n \approx 0.76$), we have not been able to resolve the corresponding  discontinuity of either magnetization or SC gap in our simulations. This allows us to classify this transition as of a weak first order type. Unconventional topological transitions that are not associated with bulk single-particle gap closure have recently sparked discussion in the context of other materials, particularly MoTe$_2$/WSe$_2$ moiré heterobilayers \cite{Li2021,Mai2024}. The discussed here $C = -2$ TSC provides a concrete correlation-assisted mechanism resulting in such behavior.

We now supplement the analysis of the topological transitions shown in Fig.~\ref{fig:Gap_mag_Fermi}(a)-(b) by relating them to the evolution of the underlying Fermi surface (FS), here obtained by setting the SC gap to zero in the effective Hamiltonian obtained variationally within the SGA scheme \cite{Spalek2022}. Figure~\ref{fig:Gap_mag_Fermi}(c) shows the FS evolution for doping levels marked by dashed vertical lines in Fig.~\ref{fig:Gap_mag_Fermi}(a)-(b), and numbered from one to four. In each case, RSOC-induced splitting of the band structure is apparent. Starting from within the $C = -4$ state on the right (FS~4) and topologically trivial ($C = 0$) SC region in between the $C = -4$ and $C = -2$ TSC, two hole pockets around the $M$ Brillouin-zone point are observed. This reflects conventional RSOC-split high-$T_c$ fermiology and indicates that the $C = -4 \rightarrow 0$ transition does not involve reconstruction of single-particle band structure. Notably, FS undergoes a Lifshitz transition across  the boundary between the $C = -2$ and $C = 0$ states [cf. FS~3 and FS~2 in Fig.~\ref{fig:Gap_mag_Fermi}(c)]. This points toward an interrelationship between electronic structure, correlation-driven first order metamagnetic transition, and SC state topology. The Fermi surface undergoes another reconstruction on the left boundary of the $C = -2$ state [cf. FS~2 and FS~1 in Fig.~\ref{fig:Gap_mag_Fermi}(c)], which we link to a weak first-order transition without a clear discontinuity of SC amplitudes and magnetization. This interplay of the effects is most pronounced in the range of intermediate correlations ($U \sim W/2$), but it requires a separate detailed analysis.

\subsection{Strong-coupling (\textit{t-J} model) limit}
\label{subsection: t-J model}

\begin{figure}
  \centering
  \includegraphics[width=1.0\linewidth]{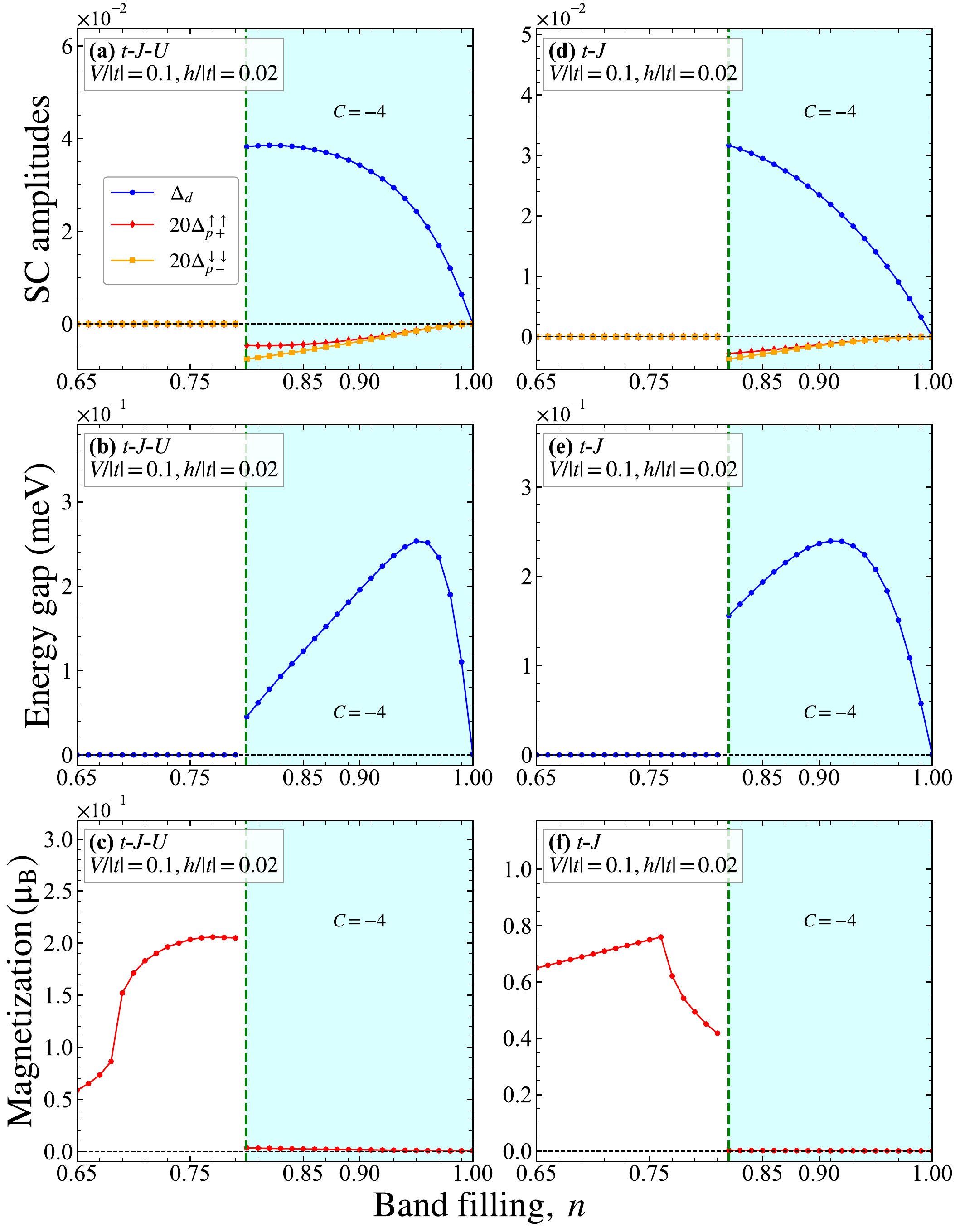}
  \caption{Comparison of the SC amplitudes [(a) and (d)], gap in the single-particle spectrum [(b) and (e)], and magnetization [(c) and (f)] between the $t$-$J$-$U$ model with on-site Coulomb repulsion $U = 14 |t|$ (left panels) and the $t$-$J$ model in the $U \rightarrow \infty$ limit (right panels). The remaining parameters are identical in all panels, $t = -0.35\,\mathrm{meV}$, $V=0.1 |t|$, $h=0.02 |t|$, $k_B T = 10^{-5} |t|$, and $N = 800 \times 800$. Green shaded areas mark the $C=-4$ TSC state. Green vertical dashed lines indicate first-order phase transition points ($n=0.799$ for $t$-$J$-$U$ and $n=0.820$ for $t$-$J$ model. Close correspondence between left- and right panels indicates that $U = 14 |t|$ case reflects already the essential features of the strong-coupling limit.}
  \label{fig:tJU_vs_tJ_3x2}
\end{figure}

We now turn to the discussion of topological phase diagram of the $t$-$J$-$U$ model~(\ref{eq:Hfinal}) in the $U \rightarrow \infty$ limit, effectively eliminating configurations involving double occupancies. The Hamiltonian then reduces to the $t$-$J$ model \cite{Chao1977}, supplemented with RSOC and Zeeman field. In Fig.~\ref{fig:tJU_vs_tJ_3x2} we compare the SC amplitudes, energy gap in the quasiparticle spectrum, and magnetization obtained within the $t$-$J$-$U$ model ($U = 14 |t|$) and $t$-$J$ model framework ($U = \infty$). The remaining parameters are the same as those adopted in Fig.~\ref{fig:Phase_D_V0.1_h0.02}, i.e. $t^\prime = 0.25 |t|$, $J = 0.3 |t|$, $V = 0.1 |t|$, $h = 0.02 |t|$, $N = 800 \times 800$, and $k_B T = 10^{-5} |t|$. To accurately resolve the thermodynamic phase boundaries, both data sets utilize identical free-energy optimization scheme (cf. Appendix~\ref{appendix:hysteresis}). While the transition point undergoes a small shift ($n = 0.799$ for $t$-$J$-$U$ and $n = 0.820$ for $t$-$J$ model, the band-filling evolution of the calculated amplitudes is qualitatively similar in both cases. In particular, a clear discontinuity of calculated physical quantities, accompanied by a transition into the $C=-4$ phase (blue shaded areas) is observed. These results demonstrate that the upper portion ($U \sim 14 |t|$) of the phase diagram presented in Fig.~\ref{fig:Phase_D_V0.1_h0.02} already captures the essential features of strong-coupling limit. In particular, no $C = -2$ TSC appears near $n \sim 0.7$-$0.8$ in the regime of strong correlations.

\section{Summary and outlook}
\label{section:summary}

In this work we have investigated the interplay between local electronic correlations (controlled by on-site Coulomb repulsion), Rashba spin-orbit coupling (RSOC), and applied Zeeman field within the framework of extended $t$-$J$-$U$ model. The latter has been analyzed using statistically-consistent Gutzwiller approach (SGA) that accounts for electronic correlation effects and is applicable to large lattices. The constructed electronic density (band filling) vs. on-site Coulomb repulsion phase diagram has been analyzed in both strong-coupling regime relevant to high-temperature copper-oxide superconductors (as motivated by spin-polarized ARPES \cite{Gotlieb2018, Iwasawa2023, Luo2024}) and intermediate-to-weak correlation range. The discussion of the full $U/|t|$ range has been possible by starting from the $t$-$J$-$U$ model, within which unconventional superconductivity emerges already on the SGA (renormalized mean-field) level. In the high-$T_c$ cuprate parameter range, we have identified a transition to a robust Chern number $C = -4$ TSC state close to half-filling, which is driven by closure of the gap in the single-particle spectrum. Electronic correlations are not instrumental for the $C = -4$ state stabilization as it appears already in Bogoliubov de Gennes scheme. However, we have observed also a distinct $C = -2$ TSC regime on hole-doped side of the phase diagram that persists up to intermediate correlation regime, and is reached via first-order concomitant metamagnetic and SC transitions. The role of electronic correlations as the driving force of this transition has been established by demonstration of discontinuous evolution of magnetization, singlet- and triplet SC amplitude components, as well as of Lifshitz transition of the RSOC-split band structure. Our work thus points toward relevance of electronic correlation effects to topological phase diagrams of RSOC-coupled superconductors.

In the present study we have limited ourselves to the solutions without long-range spin-density-wave (SDW) or charge-density-wave (CDW) orders. It is empirically established \cite{Wen2019} that both CDW and SDW proliferate within the phase diagram of copper oxide superconductors, and are also resolved theoretically within SGA \cite{Spalek2022} and related slave-boson schemes \cite{Igoshev2015}. Coupling of the SDW/CDW to TSC solutions discussed above is expected to induce more complex states than considered so far, including topological pair-density wave solutions. The construction of such realistic phase diagrams in the presence of electronic correlations is challenging due to complexity of the resulting order parameters and large supercells required to account for spatially inhomogeneous configurations. However, incorporation of antiferromagnetism and commensurate CDW remains feasible within the SGA scheme, and should be addressed in a separate study. Another natural extension of our work is construction of the correlated phase diagram analogous to that of Fig.~\ref{fig:Phase_D_V0.1_h0.02} in Zeeman field tilted away from out-of-plane direction. All of these features could provide clues for new unique phases to be looked for in experiment.

\section*{Acknowledgments}

This work was supported by Grants UMO-2021/41/B/ST3/04070 and UMO-2023/49/B/ST3/03545 from Narodowe Centrum Nauki (NCN), and by Grant ``Research Support Module'' (RSM/119/DT) as a part of the ``Excellence Initiative-Research University'' program at Jagiellonian University in Kraków. For the purpose of Open Access, the authors have applied a CC-BY public copyright licence to any Author Accepted Manuscript (AAM) version arising from this submission.

\appendix

\section{Characterization of the first-order phase transition}
\label{appendix:hysteresis}

\begin{figure}
  \centering
  \includegraphics[width=1\linewidth]{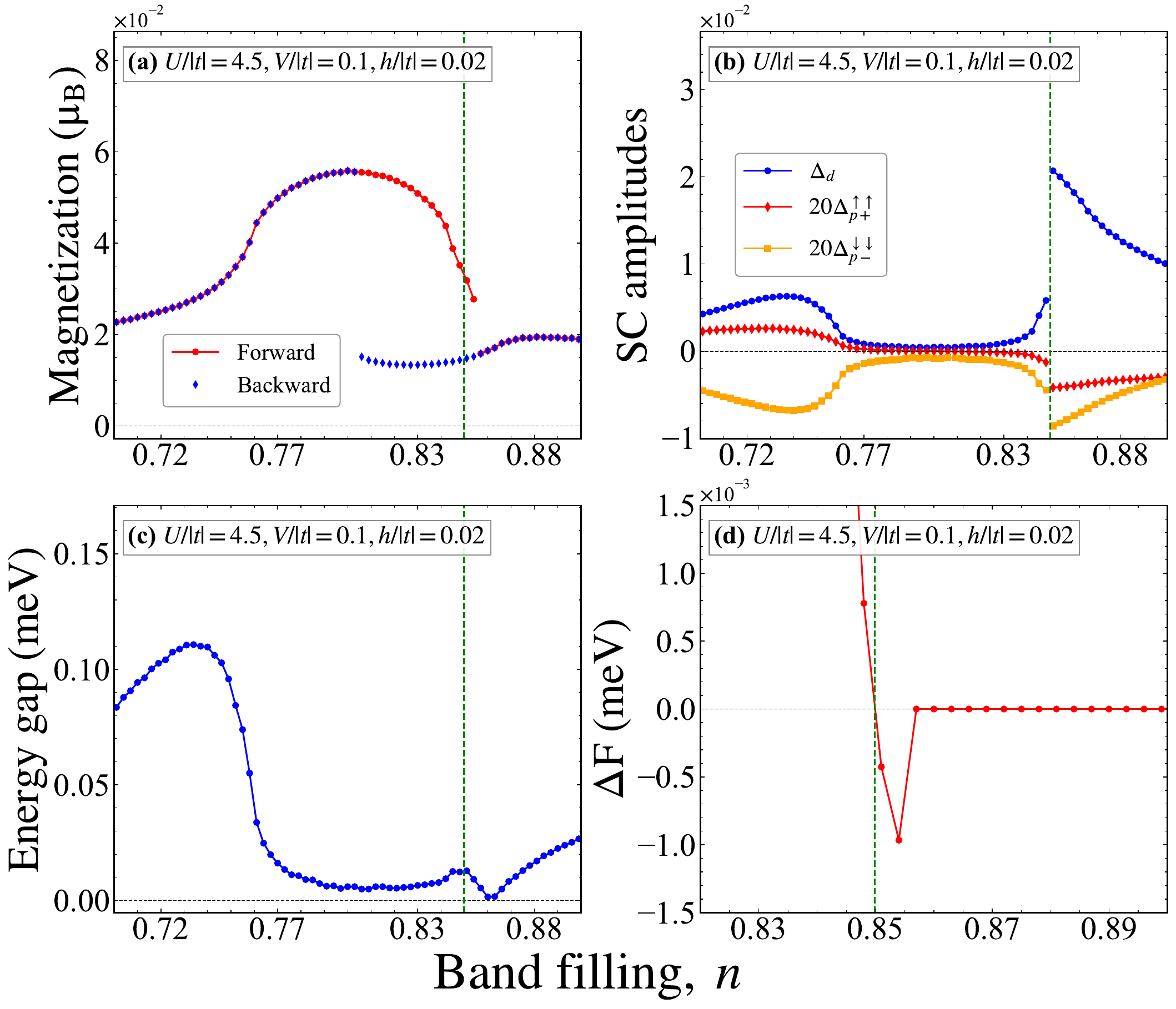}
  \caption{The construction used to determine the first-order phase transition boundary. Panel (a) shows magnetization as a function of band filling $n$, evaluated via forward (red) and backward (blue) density sweeps. A broad phase coexistence regime is observed. Panel (b) details the corresponding superconducting amplitudes (singlet $\Delta_d$ alongside scaled triplet components $20\Delta_{p+}^{\uparrow\uparrow}$ and $20\Delta_{p-}^{\downarrow\downarrow}$), exhibiting sharp discontinuity, which is not assisted by bulk single-particle energy gap closure [panel (c)]. Panel (d) details the free-energy difference $\Delta F = F_{\mathrm{backward}} - F_{\mathrm{forward}}$ between the metastable states. The vertical green dashed line across all panels denotes the true thermodynamic transition point at $\Delta F = 0$ (estimated as $n=0.849$). The model parameters are: $t = -0.35\,\mathrm{eV}$, $U=4.5 |t|$, $V=0.1 |t|$, $k_B T = 10^{-5} |t|$, $h=0.02 |t|$, and $N=800 \times 800$.}
    \label{fig:hysteresis_and_jumps}
\end{figure}

Here we analyze the first-order transition between topologically trivial and $C=-2$ TSC states. To construct the horizontal cuts in the phase diagram of Fig.~\ref{fig:Phase_D_V0.1_h0.02}, we used two data sets representing adiabatic forward and backward sweeps of the band-filling axis, $n$. Near the first-order transition point, the converged solution is sweep-dependent, indicating phase coexistence. In Fig.~\ref{fig:hysteresis_and_jumps}(a), we illustrate the magnetization hysteresis loop for the horizontal cut of the phase diagram at $U/|t|=4.5$. The remaining parameters are $t = -0.35\,\mathrm{eV}$, $t^\prime = 0.25 |t|$, $J = 0.3 |t|$, $V = 0.1 |t|$, $h = 0.02 |t|$, $k_B T = 10^{-5} |t|$, and $N = 800 \times 800$. The thermodynamic transition point cannot be deduced solely from the extent of the hysteresis loop, as its boundaries are dictated by spinodal instabilities where a given metastable branch is no longer supported. We determine critical band-filling based on the analysis of the free-energy difference between the two competing solutions, $\Delta F = F_{\text{backward}} - F_{\text{forward}}$, utilizing the construction depicted in Fig.~\ref{fig:hysteresis_and_jumps}(d). The thermodynamic phase transition occurs at $\Delta F = 0$, in this case yielding a critical band filling of $n_c = 0.849$. For $n < n_c$, the highly magnetized state tracked via the forward sweep minimizes the free-energy functional $\mathcal{F}$\cite{Fidrysiak2023}, while the low-magnetization state tracked in the backward sweep remains stable for $n > n_c$. This free-energy cross-over verification ensures that the reported phase boundaries represent true thermodynamic transitions.

For completeness, we evaluate the density-driven evolution of the superconducting order parameters [Fig.~\ref{fig:hysteresis_and_jumps}(b)] and the bulk single-particle energy gap [Fig.~\ref{fig:hysteresis_and_jumps}(c)]. The discontinuous nature of the first-order phase transition is unambiguously manifested as sharp jumps in the coexisting pairing amplitudes. Crucially, this thermodynamic instability facilitates a change in the topological invariant without a requisite closure of the bulk gap, a phenomenon previously noted in Fig.~\ref{fig:Gap_mag_Fermi}(f). This unconventional gapless topological transition is explicitly corroborated in panel (c), where the finite gap magnitude is maintained across the exact transition point marked by the vertical green dashed line.

\section{Finite-size effects}
\label{appendix:finite_size}

\begin{figure}
  \centering
  \includegraphics[width=1.0\linewidth]{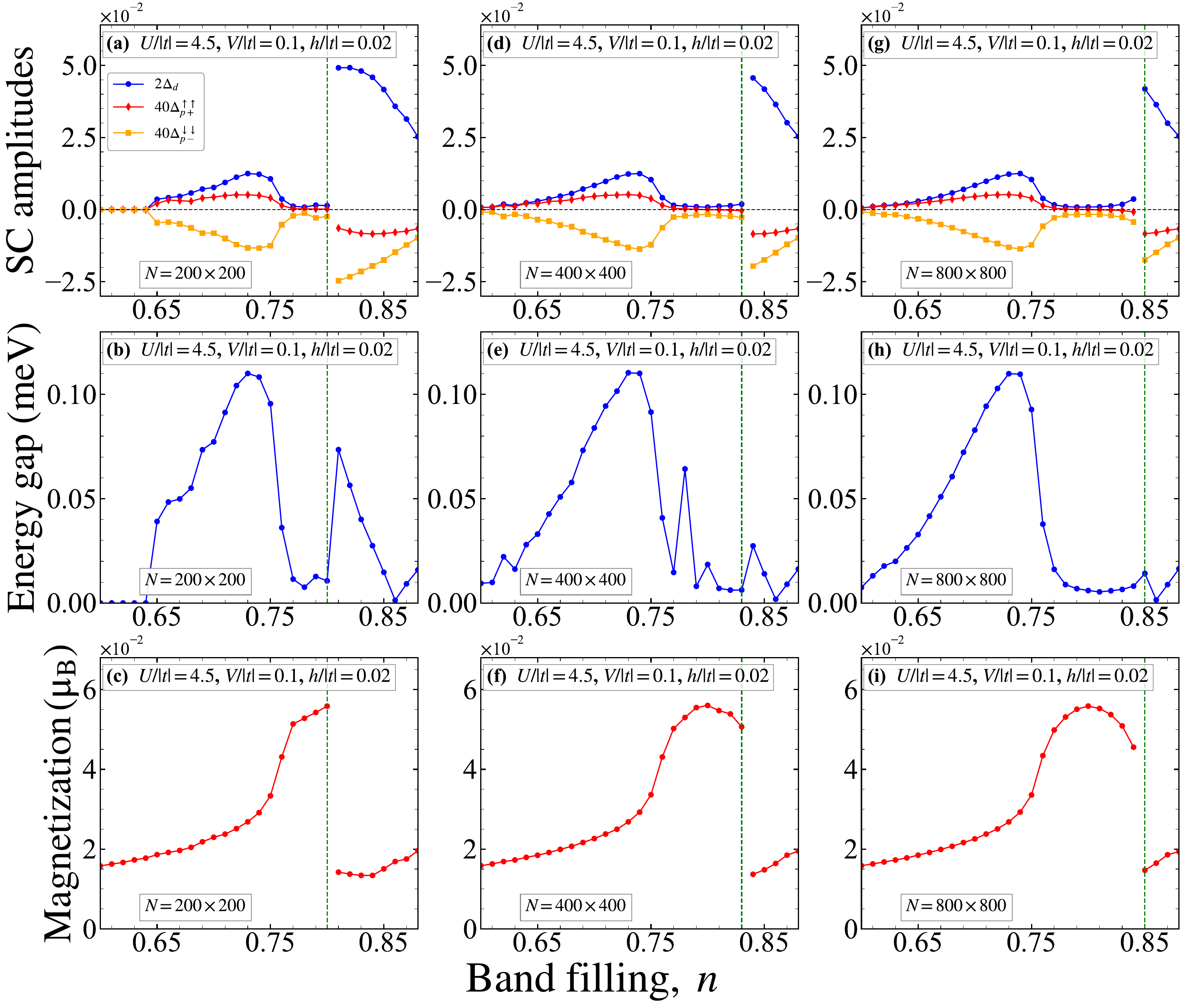}
  \caption{Finite-size scaling of the SC amplitudes (top panels), bulk single-particle energy gap (middle panels), and magnetization (bottom panels). All quantities are plotted as a function of the band filling, $n$. The adopted lattice sized are (a)-(c) $N = 200 \times 200$, (d)-(f) $N = 400 \times 400$, and (g)-(i) $N = 800 \times 800$. Comparison between the columns demonstrates that increasing the system size leads to systematic reduction of the finite-size effects, with $800 \times 800$ lattice being sufficient to account for small-gap SC close to the metamagnetic transition. The model parameters are: $t = -0.35\,\mathrm{eV}$, $t^\prime = 0.25 |t|$, $J = 0.3 |t|$, $U = 4.5 |t|$, $V = 0.1 |t|$, $h = 0.02 |t|$, and $k_B T = 10^{-5} |t|$.}
  \label{fig:finite_size_effect}
\end{figure}

In the presence of RSOC and Zeeman field, small gaps open at the nodes in single-particle spectrum, resulting in emergence of strong TSC for appropriately selected model parameters (cf. Sec.~\ref{section:results}). Simulations of TSC carried out for finite lattices are thus prone to finite-size effects whenever mean energy difference between discrete energy levels approach the SC gap magnitude. Here we discuss these aspects and assess the convergence of our numerical results to the thermodynamic limit as the lattice size $N$ increases. The employed RSOC and Zeeman field coincide with those adopted in Sec.~\ref{section:results}, i.e., $V = 0.1 |t|$ and $h=0.02 |t|$. The remaining parameters are $t = -0.35\,\mathrm{eV}$, $t^\prime = 0.25 |t|$, $J = 0.3 |t|$, $U = 4.5 |t|$, and $k_B T = 10^{-5} |t|$.
 
Figure~\ref{fig:finite_size_effect} shows calculated band-filling dependence of the SC amplitudes (top panels), energy gap in the quasiparticle spectrum (middle panels), and magnetization (lower panels) for lattice size $N = 200 \times 200$ [(a)-(c)], $N = 400 \times 400$ [(d)-(f)], and $N = 800 \times 800$ [(g)-(i)]. The choice of a $200 \times 200$ lattice results in sizable fluctuations of the SC gap magnitude, cf. Fig.~\ref{fig:finite_size_effect}. This may be interpreted by estimating an average distance between energy levels $\Delta E(N) \sim \frac{8|t|}{N}$, where $8|t|$ is a bare single-particle bandwidth. We find that $\Delta E(200 \times 200) \approx 0.07\,\mathrm{meV}$, which is comparable to the maximal value of SC gap $\sim 0.1\,\mathrm{meV}$ seen in Fig.~\ref{fig:finite_size_effect}(b). For larger systems we obtain $\Delta E(400 \times 400) \approx 0.018\,\mathrm{meV}$ and $\Delta E(800 \times 800) \approx 0.0044\,\mathrm{meV}$. This demonstrates that $800 \times 800$ lattice is sufficient to account for the small-gap ($\sim 0.02\,\mathrm{meV}$) SC state close to the first-order phase transition at $n \approx 0.85$ (cf. the magnetization discontinuity in bottom panels). 
 
%

\end{document}